\documentclass[12pt]{iopart}
\usepackage{graphicx}
\usepackage[latin1]{inputenc}
\usepackage[dvips]{color}
\begin{document}

\title[Curved Graphene Nanoribbons: Structure and Dynamics of Carbon Nanobelts]{Curved Graphene Nanoribbons: Structure and Dynamics of Carbon Nanobelts}
\author{B V C Martins$^1$ , D S Galv\~ao$^1$}
\address{$^1$ Instituto  de Física ''Gleb Wataghin'', Universidade Estadual de Campinas, Unicamp 13083-970, Campinas, São Paulo, Brazil}
\ead{galvao@ifi.unicamp.br}

\begin{abstract}

Carbon nanoribbons (CNRs) are graphene (planar) structures with large aspect ratio. Carbon nanobelts (CNBs) are small graphene nanoribbons rolled up into spiral-like structures, i. e., carbon nanoscrolls (CNSs) with large aspect ratio. In this work we investigated the energetics and dynamical aspects of CNBs formed from rolling up CNRs. We have carried out molecular dynamics simulations using reactive empirical bond-order potentials. Our results show that similarly to CNSs, CNBs formation is dominated by two major energy contribution, the increase in the elastic energy due to the bending of the initial planar configuration (decreasing structural stability) and the energetic gain due to van der Waals interactions of the overlapping surface of the rolled layers (increasing structural stability). Beyond a critical diameter value these scrolled structures can be even more stable (in terms of energy) than their equivalent planar configurations. In contrast to CNSs that require energy assisted processes (sonication, chemical reactions, etc.) to be formed, CNBs can be spontaneously formed from low temperature driven processes. Long CNBs (length of $\sim$ 30.0 nm) tend to exhibit self-folded racket-like conformations with formation dynamics very similar to the one observed for long carbon nanotubes. Shorter CNBs will be more likely to form perfect scrolled structures. Possible synthetic routes to fabricate CNBs from graphene membranes are also addressed.
\end{abstract}
\pacs{6148 Gh}
\maketitle

\section{Introduction}

\begin{figure}[ht!] \centering
\includegraphics[width=14cm]{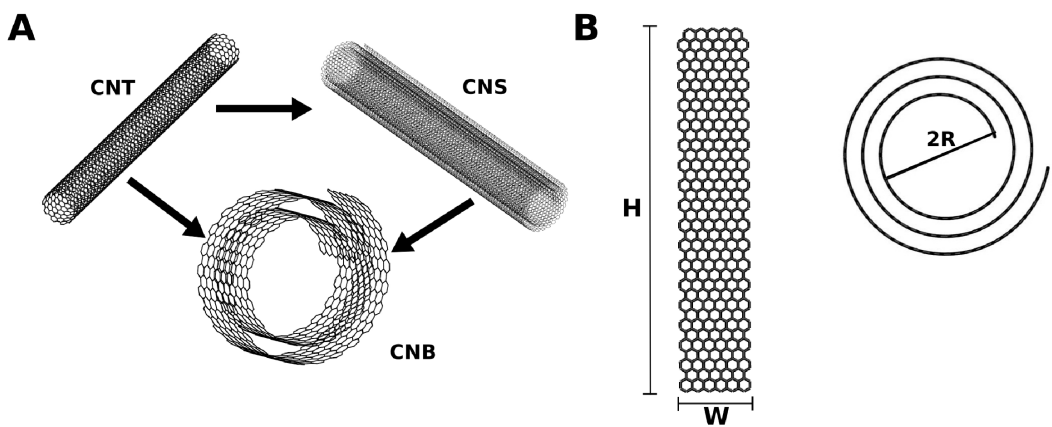}
\caption{ A: Carbon Nanotube (CNT), Carbon Nanoscroll (CNS) and Carbon Nanobelt (CNB). See text for discussions; B: Belt dimensions: Height H, Widht W and the approximated internal diameter (2R).}
\label{fig:schematic}
\end{figure}

\indent The discovery of carbon nanotubes (CNTs) resulted in a revolution on the research of carbon-based materials \cite{iijima}. Despite of their large number of non-usual mechanical and electronic properties, the nanotube structure is very simple. It is topologically equivalent to the rolling up of one or more graphene sheets into seamless cylinders.  However, CNT cap closure results in a rigid closed structure with no possibility of significantly altering its internal diameter. Such a dynamical property would be highly desirable for structural adaptation to different molecular intercalants, such as, hydrogen for energy storage \cite{galvao1}. This is one of the motivations for the research of the so-called carbon nanoscrolls (CNSs) \cite{iijima,galvao1,tomanek,galvao2,viculis,akita,jiang,perim}, which consists of rolled spiral-like graphene sheet or sheets (Fig. \ref{fig:schematic}). Similarly to CNTs, single and multisheets CNSs are possible. The thermodynamics and structural stability of CNTs and CNSs have different origins, the CNS stability being the result of the balance between the elastic tension of rolling up graphene sheets (negative contribution) and van der Waals interactions among adjacent sheet layers (positive contributions). The scroll-like topology allows an easy internal volume modification that can be the basis of a series of applications that are not possible with nanotubes \cite{galvao2,rurali}.

\indent Large carbon nanoscroll structures have been synthesized \cite{viculis,akita} based on a procedure starting from the exfoliation of highly oriented pyrolytic graphite (HOPG) crystals using KOH as intercalant, followed by sonication at high temperatures ($250^oC$). It was showed that this procedure can break the graphene sheets into small flakes of nanometer size. More recently \cite{jiang} a new, simple and effective way of fabricating high-quality CNSs from graphene structures deposited over Si substrates has been achieved. Depending on the aspect ratio it is possible to create the so called carbon nanobelts (CNBs) or carbon nanoribbons (CNRs). CNRs consists of small graphene stripes with high aspect ratio (length/width). They can present some unusual electronic and mechanical properties \cite{hod,louie,querlioz,fertig,ezawa,white,zhang,faccio,qli,schniepp,yu,cantele,geim} making them an alternative to nanotubes in some applications \cite{guo,chen,cohen,stettler}. Their experimental realization, reported on a recent work \cite{wang}, followed similar procedures to the route of synthesis of carbon nanoscrolls. Theoretical studies about saddle-shape nanoribbons were published recently \cite{yakobson,zhang,galvao3,thesis}. There are also some studies about folding and/or structural collapse carbon nanotubes \cite{buehler,buehler2}. The CNR structural similarity with nanoscrolls suggests that nanobelts (nanoribbons rolled up into scroll-like structure) can also have stable conformations (fig. \ref{fig:schematic}). 

\indent In this work we have theoretically investigated the structural and dynamics aspects of nanobelts formation from graphene nanoribbons. We carried out molecular mechanics and molecular dynamics simulations of the process of rolling up nanoribbons starting from planar conformations.

\section{Methodology}

\indent The formation of belt/spiral structures starting from graphene planar structures is a continuous dynamical process since the equilibrium condition (stable or meta-stable) is determined by the balance of simultaneous  repulsive elastic tension and attractive van der Waals interactions. As a typical nanobelt structure can have more then 1000 atoms and we need to analyze many structures of different sizes and at different temperatures, the use of sophisticated ab initio methods is cost prohibitive. We then choose to use classical molecular mechanics and molecular dynamics methods. As the van der Waals interactions are of fundamental importance to determine the scrolling dynamics we used a modified reactive empirical bond-order Brenner potential \cite{brenner,susan} where the van der Waals interactions were included using a Lennard-Jones function. In their original form the Brenner potential did not include these interactions. For details see \cite{galvao4} and references cited therein. The use of such potentials is important for cases where bond breaking and/or new bonds formation can occur. This is not the case in standard molecular force fields (that often use quadratic form terms which can not handle these cases). In our present study it is important to consider the possibility of breaking/formation of bonds since their  formation connecting different scrolled layers would prevent the scrolling process of non-passivated structures. This kind of reactive potential has been proven to be accurate in describing carbon-based $sp^2$ structure deformations \cite{galvao4}.
       
\indent For molecular mechanics analysis we considered single point energy (static) calculation of specific configurations. For molecular dynamics analysis we ran simulations for the necessary time to obtain equilibrated configurations. We used a Berendsen thermostat \cite{berendsen} and a predictor-corrector integration algorithm with time steps of 1 fs. All dynamical simulations followed a heating ramp starting from 10 K until 100 K with 10 K variation and 50 ps step length.

\indent The model structures were generated by cutting a graphene sheet along the armchair axis (Fig. 1). We used nanoribbons with width (W) of 1.2 nm (5 aromatic rings)  and lengths (L) varying from approximately 10.0 up to 30.0 nm (nominal values). This width was chosen based on the fact that this value is commonly found in the experiments producing graphene nanoribbons \cite{wang}. For each (W,L) dimension pair value we constructed a passivated and a  non-passivated belt. We have included the non-passivated structures in our analysis in order to determine the role of bond formation on the scrolling processes. The initial planar conformation was randomly perturbed by small (0.1 nm) deviations on the plane axis for the molecular dynamics simulations to avoid local minima.

\indent In this work we decided to restrict ourselves to present only the results to armchair-edged nanobelts by two main motives. Firstly, from our preliminary investigations we observed that the scrolling processes would be easier to armchair in comparison to zig-zag ones. Secondly, and more important, recent experiments from structural graphene nanoribbon reconstructions \cite{girit} indicated that zigzag edged structures are much more stable than armchair ones.

\section{Results and Discussion}

\indent One of the main objectives of the present study is to verify whether and at what circumstances CNBs can be spontaneously formed from CNRs upon heat treatment. In principle, depending on geometrical factors, it is possible that scrolled configurations (CNBs) can be more stable (in terms of energy) than their equivalent planar configurations \cite{galvao1}.

\begin{figure}[ht!] \centering
\includegraphics[width=10cm]{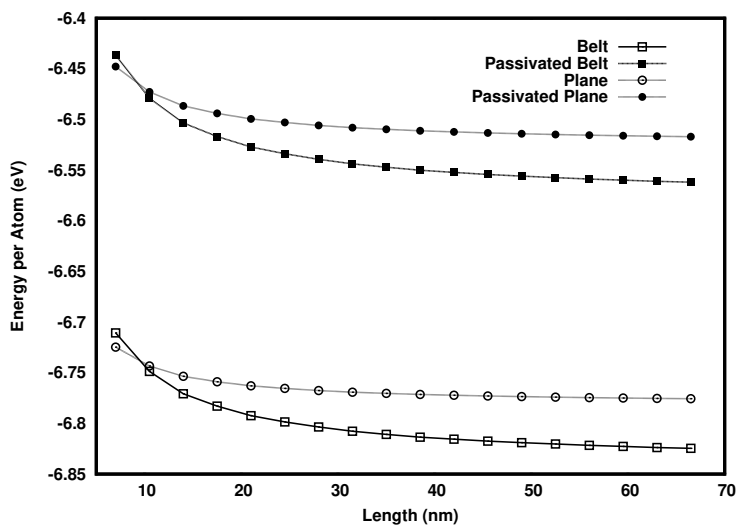}
\caption {Single point energy calculations as a function of length for a fixed ribbon 
width (1.2 nm) in planar and belt configurations.}
\label{fig:energy}
\end{figure}

In Fig. \ref{fig:energy}  we present the energy per atom (static calculations) for planar and scrolled configurations as a function of H (see Fig. 1) for structures with a fixed W=1.2 nm. We have considered passivated and non-passivated structures. The results are for scrolled structures with internal diameter about 25 $\AA$ and rolled up into an Archimedean spiral with interlayer distances of $\sim$ 3.4 $\AA$.  These numbers were obtained following the same simulation protocols described in refs. \cite{galvao1,perim}. 

As we can see from Fig. \ref{fig:energy}, after a critical length of H $\sim$ 10.0 nm all scrolled configurations are more stable than their equivalent planar ones. The explanation for this behavior is the same for larger scrolls \cite{galvao1,perim}. The scrolling process, starting from a planar configuration, involves two major energy contributions; the elastic energy increase caused by bending the layer (decreasing stability) and the free energy decrease generated by the van der Waals interactions (increasing stability) of the overlapping layers. As the structure is rolled up and before layer overlap occurs, the increase in curvature decreases structural stability due to the increase in the elastic energy. Thus, the structure becomes less stable in relation to its undistorted (planar) configuration. If the structure is allowed to relax it will return to its more stable planar configuration. However, if the rolled layers overlap the van der Waals energy increase stability and the final configuration will be determined by the geometrical features. For small diameter values the bending contributions outweigh the van der Waals gains and the scrolled structure is unstable and tends to evolve into planar structures. After a critical inner diameter value where the van der Waals gains outweigh the losses in stability by the elastic deformation stable scrolled can be formed and some of them can be even more stable (in terms of energy) than their planar configurations, as for the cases indicated in Fig. \ref{fig:energy}.

The high stability of the planar configurations and the existence of a critical diameter value to stabilize the scrolled configuration implies that the transition from planar to rolled structures must be energy assisted, e. g., through sonication \cite{viculis,akita} or exothermic chemical reactions \cite{jiang}, as recently experimentally reported. For the CNBs, due to their small dimensions it is possible that the rolling processes could be thermally driven. In order to test this hypothesis we used molecular dynamics simulations.

The possibility of CNB thermally driven spontaneous formation was tested using a typical passivated armchair nanoribbon (H=30.0 nm, W=1.2 nm, see Fig. \ref{fig:schematic}). The MD simulations were carried out using a heat ramp, starting from 10 K up to 100 K. The temperature is increased in steps of 10 K, with runs of 50 ps for each temperature step. After reaching the target temperature of 100 K, an extra MD run of 50 ps was carried out in order to assure high quality thermal equilibration.

\begin{figure}[ht!] \centering
\includegraphics[width=12cm]{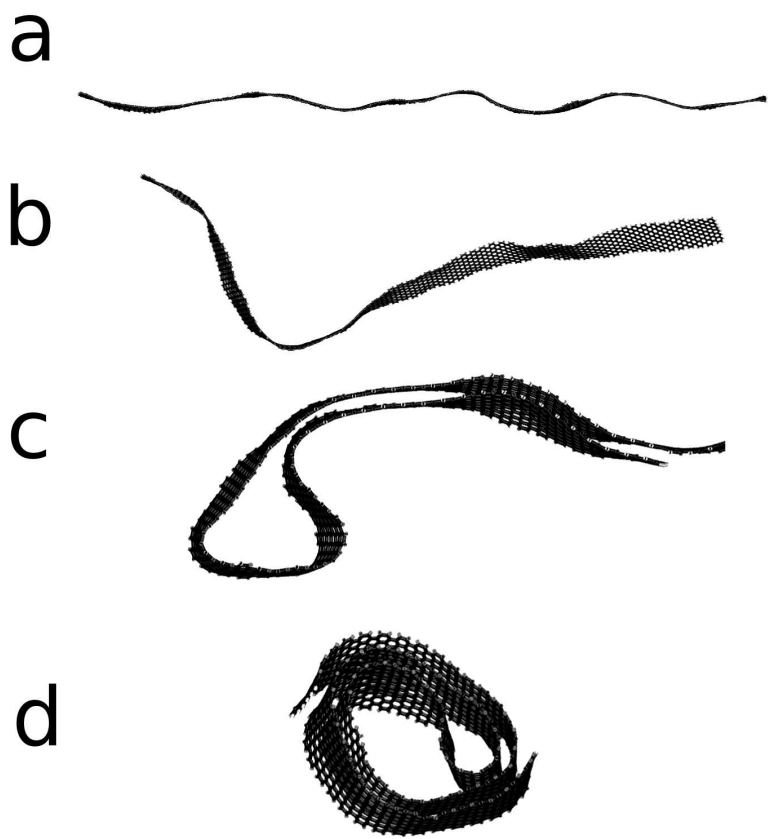}
\caption{Representative snapshots from molecular dynamics simulations for a 30 nm belt: a) 23.7 ps; b) 298 ps; c) 380.5 ps; d) 500 ps}
\label{fig:force}
\end{figure}

\begin{figure}[ht!] \centering
\includegraphics[width=12cm]{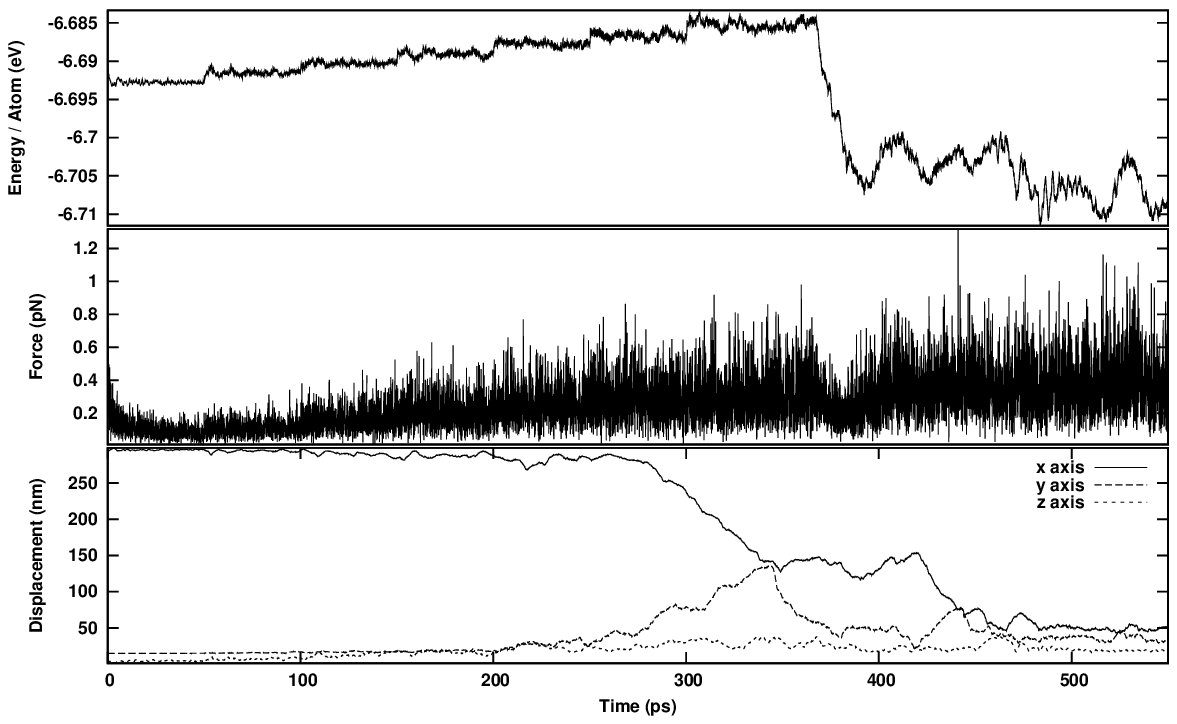}
\caption{Top: Potential energy / atom; middle: force modulus; bottom: Mean value
for the coordinate projections on the principal axes of the structure}
\label{fig:plots}
\end{figure}

In Fig. \ref{fig:force} we present representative snapshots from the MD simulations. The corresponding time evolution profiles for energy, force and positions are presented in Fig. \ref{fig:plots}.

Our results showed that the thermal energy induced the appearance of border oscillations (Fig. \ref{fig:force}a). These oscillatory patterns have been recently reported for related structures \cite{zhang,yakobson} and it is believed to be created by edge stresses. Amplified by thermal energy these oscillations can bend and twist the nanoribbons (Fig. \ref{fig:force}b). If the surfaces come closer enough the van der Waals interactions can lead to the formation of racket-like self-folded nanoribbons (Fig. \ref{fig:force}c). Once formed the large contact surface area produces a significant van der Waals energy gain (significant drop in the potential energy in Fig. \ref{fig:plots}). Depending on the temperature and geometry, this gain can not be overcome by the thermal energy and the unfolding can not occur. These folded structures do not have anymore significant border oscillations, but this not prevent the bending processes to continue and the structures can evolve towards rolled ones (Fig. \ref{fig:force}d). However, the thermally driven self-folded configurations at the early stages of the dynamics processes prevents the formation of truly CNBs (perfect scroll-like Archimedean spirals). These processes can be better visualized in the video of the supplementary materials. Interestingly, this observed self-folding dynamics is very similar to the one reported to long CNTs \cite{buehler,buehler2}. This phenomenon is again the result of an interplay between bending stiffness and binding energies, which balance is strongly aspect ratio dependent.

\begin{figure}[ht!] \centering
\includegraphics[width=10cm]{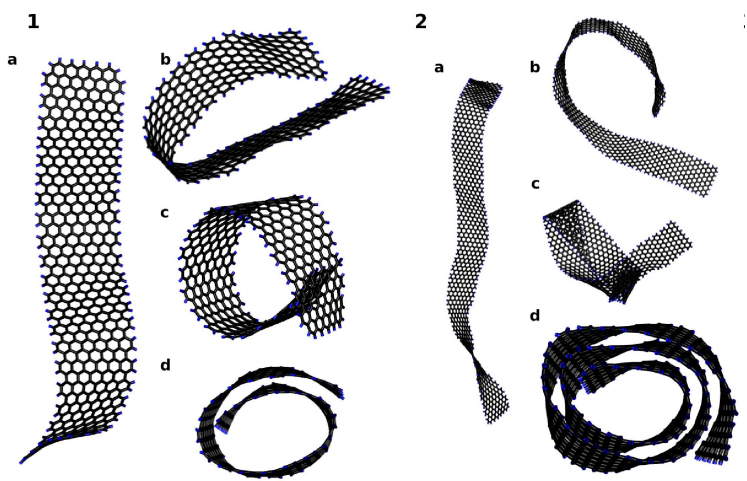}
\caption{Representative snapshots from molecular dynamics simulations for passivated structures. Set 1 (L=10 nm): a) 350 ps; b) 600 ps; c) 625 ps; d) 650 ps; Set 2 (L=20 nm): a) 350 ps; b) 400 ps; c) 412 ps; d) 500ps}
\label{fig:others}
\end{figure}

Based on that we have investigated whether varying the CNR aspect ratio (varying the length, but keeping the same width, see Fig.\ref{fig:schematic}) could increase the chances of producing perfect CNBs. In order to test this hypothesis we have extensively investigated (using the same simulation conditions) CNRs of different sizes. We observed that the chances of forming perfect CNBs are increased decreasing the nanoribbon length. In Fig. \ref{fig:others} we present representative snapshots from the MD simulations. We observed CNB formation for structures of lengths of 10 and 20 nm. Decreasing the CNR length alters the relative contribution of energy bending and van der Waals energies. For smaller structures the thermal energy fluctuations are more likely to occur and this makes more difficult the formation of permanent racket-like self-folded structures, thus increasing the chances of forming perfect CNBs (see video in supplementary materials).

These results show that although CNSs and CNBs share some of the fundamental aspects of their dynamics of formation, due to their smaller dimensions (in comparison with CNSs) CNBs present some peculiar features. Both scrolled structures are the result of the competing processes of bending deformation and wan der Waals interactions. However, for typical CNSs \cite{viculis,akita,jiang} their formation require energies larger than the ones than can be provided by low temperature processes. In contrast, as we showed, CNBs can be spontaneously formed just heating up the structures at low temperatures. Also, while for CNBs the edge stresses and twistings play a fundamental role in the scrolling dynamics, they are of secondary importance in the CNSs formation \cite{galvao1}.

Another aspect is the relative importance of the border passivation on the spiral formation. For large structures like CNSs it is possible to break/heal new bonds formed from dangling bonds, as experimentally observed for the case of concentric CNTs \cite{zettl}. For the CNBs at low temperatures the thermal energy could not be enough to break/heal these bonds, which consequently could prevent or block the scrolling processes. 

In order to illustrate these aspects we present in Fig. 6 snapshots from MD simulations for non-passivated structures (lengths of 10.0, 20.0 and 30.0 nm) obtained using the same simulation protocols for the cases of passivated structures. Our results show that (see also video in the supplementary materials), although the configurations are quite similar to the ones obtained from passivated structures, when a few number of dangling bonds create new bonds connecting different layers, this is enough to block or prevent the scrolling process. When a significant number of these bonds are present a glassy-carbon structures can be formed. These results suggest that although the recent experimental techniques developed to produce high quality CNSs from graphene membranes \cite{jiang} could be easily adapted to fabricate CNBs, the number of dangling bonds (defects), in contrast to the CNSs, could be a limiting efficiency factor. 

\begin{figure}[ht!] \centering
\includegraphics[width=14cm]{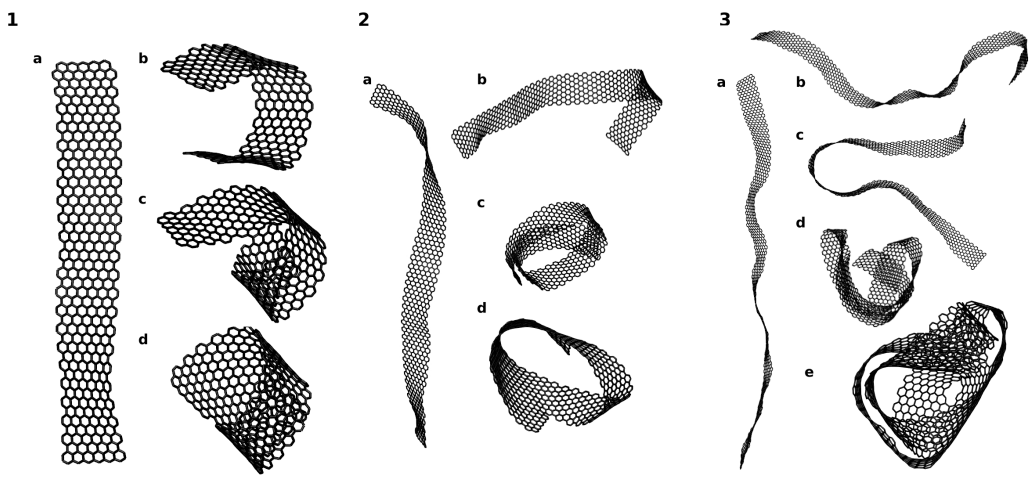}
\caption{Results for the three non-passivated simulations. Set 1 (L=10 nm): a) 150 ps; b)350 ps; c) 360 ps; d) 500 ps; Set 2 (L=20 nm): a) 350 ps; b) 400 ps; c) 412 ps; d) 500 ps; Set 3 (L=30 nm): a) 200 ps; b) 300 ps; c) 350 ps; d) 450 ps; e) 500 ps}
\label{fig:non-passivated}
\end{figure}

\section{Summary and Conclusions}
	
\indent We have investigated the energetics and dynamical aspects of carbon nanobelts (CNBs) formed from rolling up carbon nanoribbons (CNRs). CNBs are small CNRs rolled up into spiral-like structures, i. e., nanoscroll-like structures with large aspect ratio (see Fig. 1). We have carried out molecular dynamics simulations with reactive empirical bond-order potential \cite{brenner,susan}.

Our results show that the CNB formation mechanisms are quite similar to those of typical large carbon and boron nitride nanoscrolls \cite{galvao1,perim}. Similarly to carbon nanoscrolls (CNSs), CNBs formation is dominated by two major energy contributions, the increase in the elastic energy due to the bending of the initial planar structure (decreasing structural stability) and the energetic gain due to van der Waals interactions of the overlapping surfaces of the rolled layers (increasing structural stability). For both CNSs and CNBs there is a critical diameter value for rolled structure stability, beyond this limit the structures can be even more stable (in terms of energy) than its parent planar structures \cite{galvao1}. Below this limit they undergo a transition from scrolled to their planar parent structures. 

However, in contrast to CNSs, that require energy assisted mechanisms (sonication \cite{viculis,akita} or chemical reactions \cite{jiang}, for example) to their formation, CNBs due to their smaller dimensions (in comparison with typical experimentally fabricated CNSs) can be spontaneously formed through low temperature driven processes, as demonstrated from our molecular dynamics results. Also, because of these reduced dimensions thermally induced edge stresses and twistings mechanisms play a fundamental role in the CNBs scrolling processes, while they are of secondary importance for CNS case \cite{galvao1}. Our results also showed that long CNBs (length $~$ 30.0 nm) tend to exhibit self-folded racket-like conformations, with a dynamics very similar to the one observed for long carbon nanotubes \cite{buehler}. Shorter CNBs will be more likely to form perfect scrolled structures. The recently developed experimental techniques to fabricate high quality CNSs from graphene monolayers \cite{jiang} could be easily adapted to fabricate CNBs. We hope the present work will stimulate further works along these lines.

\ack{ The authors wish to thank the Brazilian Agencies FAPESP, CAPES and CNPq for partial financial support. }

\section*{References}
\bibliographystyle{unsrt}

\end{document}